\documentclass[a4paper,fleqn]{cas-dc}

\usepackage[numbers]{natbib}  

\usepackage[switch]{lineno}  
\usepackage{adjustbox}       
\usepackage{tabularx}        
\usepackage{multirow}        
\usepackage{caption}         
\usepackage{subcaption}      
\usepackage{svg}             
\usepackage{amsmath,amssymb,amsfonts}  
\usepackage{graphicx}        
\usepackage{makecell}        
\usepackage{booktabs}        
\usepackage{float}           
\usepackage{listings}        
\usepackage{color}           
\usepackage{pifont}          
\usepackage{xcolor}          
\usepackage{pdflscape}       
\usepackage{afterpage}       
\usepackage{capt-of}         
\usepackage{comment}         
\usepackage{tikz,colortbl}   
\usetikzlibrary{calc}        
\usepackage{zref-savepos}    
\usepackage{bbding}           

\usepackage{hyperref}
\hypersetup{
    colorlinks=true,
    linkcolor=blue,
    citecolor=blue,
    urlcolor=blue
}

\newcounter{NoTableEntry}
\renewcommand*{\theNoTableEntry}{NTE-\the\value{NoTableEntry}}

\usepackage{amsthm}  
\newtheoremstyle{bolddef}  
{\topsep}  
{\topsep}  
{\itshape}  
{}  
{\bfseries}  
{.}  
{ }  
{}  

\theoremstyle{bolddef}
\newtheorem{definition}{Definition}

\usepackage[linesnumbered,ruled,vlined]{algorithm2e}

\usepackage{academicons}
\usepackage{xcolor}

\def\BibTeX{{\rm B\kern-.05em{\sc i\kern-.025em b}\kern-.08em T\kern-.1667em\lower.7ex\hbox{E}\kern-.125emX}}

\SetKwInput{KwInput}{Input}
\SetKwInput{KwOutput}{Output}
\SetKwInput{KwFunction}{Function}

\definecolor{Gray}{gray}{0.85}
\definecolor{LightCyan}{rgb}{0.88,1,1}

\begin{document}

\let\WriteBookmarks\relax
\def\floatpagepagefraction{1}
\def\textpagefraction{.001}

\shorttitle{Managing Federated Learning on Decentralized Infrastructures as a Reputation-based Collaborative Workflow}

\shortauthors{Yuandou Wang et~al.}


\title [mode = title]{Managing Federated Learning on Decentralized Infrastructures as a Reputation-based Collaborative Workflow}

\author[1]{Yuandou Wang}[
                        orcid=0000-0003-4694-9572]
\ead{y.wang8@uva.nl}

\ead[url]{https://www.linkedin.com/in/yuandou-w-aa717b135/}

\credit{Conceptualization, Methodology, Investigation, Writing - Original draft preparation, Writing - Review \& Editing}

\affiliation[1]{organization={Multiscale Networked System, University of Amsterdam},
    addressline={Science Park 900}, 
    city={Amsterdam},
    postcode={1098 XH}, 
    country={The Netherlands}}

\author[1,2]{Zhiming Zhao}[%
   ]
\ead{z.zhao@uva.nl}
\ead[URL]{https://staff.fnwi.uva.nl/z.zhao/}

\credit{Conceptualization, Writing - Review \& Editing, Supervision, Project administration, Funding acquisition}

\affiliation[2]{organization={LifeWatch ERIC Virtual Lab and Innovation Center (VLIC)},
    addressline={Science Park 904}, 
    city={Amsterdam},
    postcode={1098 XH}, 
    country={The Netherlands}}


\begin{abstract}
Federated Learning (FL) has recently emerged as a collaborative learning paradigm that can train a global model among distributed participants without raw data exchange to satisfy varying requirements. However, there remain several challenges in managing FL in a decentralized environment, where potential candidates exhibit varying motivation levels and reliability in the FL process management: 1) reconfiguring and automating diverse FL workflows are challenging, 2) difficulty in incentivizing potential candidates with high-quality data and high-performance computing to join the FL, and 3) difficulty in ensuring reliable system operations, which may be vulnerable to various malicious attacks from FL participants. To address these challenges, we focus on the workflow-based methods to automate diverse FL pipelines and propose a novel approach to facilitate reliable FL system operations with robust mechanism design and blockchain technology by considering a contribution model, fair committee selection, dynamic reputation updates, reward and penalty methods, and contract theory. Moreover, we study the optimality of contracts to guide the design and implementation of smart contracts that can be deployed in blockchain networks. We perform theoretical analysis and conduct extensive simulation experiments to validate the proposed approach. The results show that our incentive mechanisms are feasible and can achieve fairness in reward allocation in unreliable environment settings.    
\end{abstract}


\maketitle

\section{Introduction}\label{sec:flops-intro}
Federated learning (FL) is a promising distributed machine learning (ML) paradigm that enables collaborative learning over decentralized data to mitigate many systematic privacy risks and costs resulting from traditional, centralized learning. In recent years, FL has been studied in many research and application domains, particularly in the fields of digital medicine and health~\citep{antunes2022federated}, open banking~\citep{long2020federated}, and the Internet of Things~\citep{nguyen2021federated}, in which data cannot be shared or exchanged due to privacy and security concerns. 

\begin{figure*}[!htb]
    \centering
    \includegraphics[width=\linewidth]{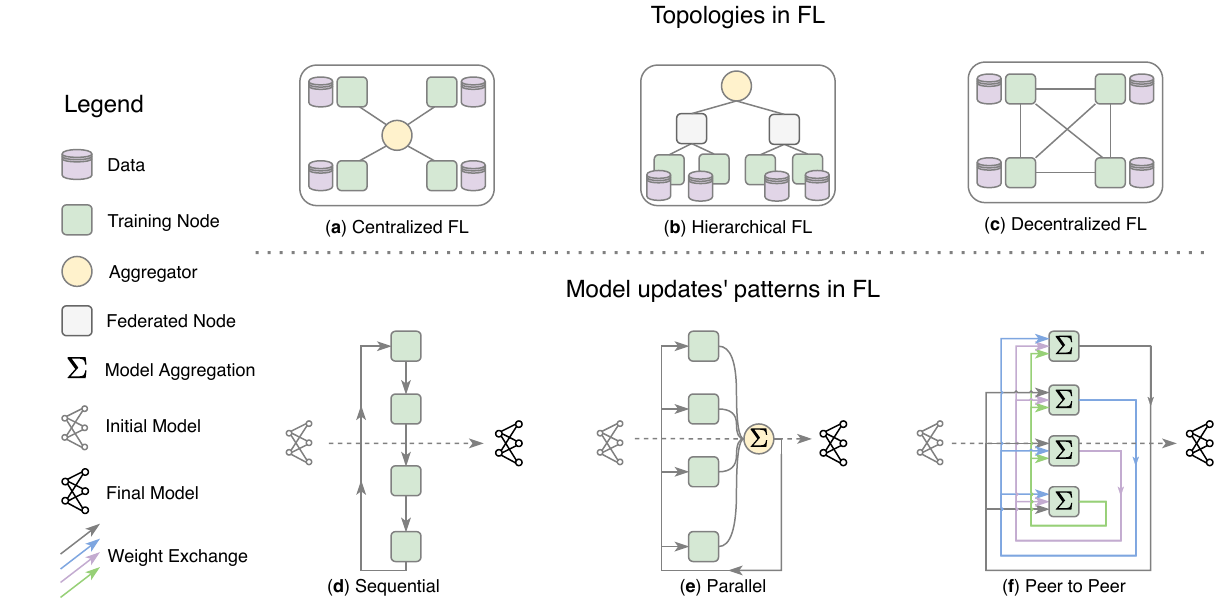}
    \caption{\textbf{An overview of diverse FL design choices.} FL topologies. (\textbf{a}) Centralized FL: a central aggregation server manages the training process by coordinating iterative training rounds. (\textbf{b}) Hierarchical FL: typically, the FL network has a tree structure with at least three tiers. (\textbf{c}) Decentralized FL: each training node is connected to one or more peers and aggregation happens on the selected node. FL model updates' paths. (\textbf{d}) Sequential. (\textbf{e}) Parallel. (\textbf{f}) Peer to Peer.}
    \label{fig:paradigm}
\end{figure*}

FL often involves a number of participants, which feature highly heterogeneous training infrastructures and non-independently and identically distributed (non-IID) data~\citep{yan2024fedrfq}. In a traditional setting of FL, several participants, each keeping its data and training process on-premise, aim to collaboratively train a joint model under a central aggregator that initializes, collects, aggregates, and redistributes the models to and from the training participants, as shown in Figure~\ref{fig:paradigm} (\textbf{a}). 
Over time, the federated model aggregation has emerged in varying modes~\citep{qi2023model} and performed as different topology types in the distributed computing ecosystem~\citep{wu2024topology}, while the data and its training infrastructure are still decentralized. It is known that such diversity can be attributed to different FL design choices and customizations to satisfy different user requirements. For instance, some researchers introduced hierarchical FL, as shown in Figure~\ref{fig:paradigm} (\textbf{b}), to tackle the bottleneck of communication overhead in the core network~\citep{xu2021adaptive, liu2022hierarchical}. Some proposed decentralized FL, as depicted in Figure~\ref{fig:paradigm} (\textbf{c}), with a Peer-to-Peer (P2P) communication architecture to cope with the drawbacks of a single point of failure and scaling issues for the increasing network size~\citep{savazzi2020federated, liu2020fedcoin}. Moreover, various aggregation strategies associated with FL topologies for updating the FL model parameters have been studied, including but not limited to sequential updates~\citep{rieke2020future}, (a)synchronous parallel updates~\citep{liu2021adaptive}, and peer-to-peer approaches~\citep{xu2022spdl}, as illustrated in Figure~\ref{fig:paradigm} (\textbf{d}), (\textbf{e}), and (\textbf{f}), respectively. The final well-trained model can be obtained through the iteratively collaborative training rounds. However, due to the heterogeneity of decentralized infrastructure or platforms, it poses execution challenges. 

FL development involves three typical activities: 1) discover potential participants with high-quality data or resource providers and stimulate them to join the FL, 2) customize the FL application workflows for specific purposes, and 3) define a reliable training process with local updates evaluation. 

In real-world scenarios, collaborative FL typically follows two main approaches. The first involves pre-established collaborations, where participants are obligated to join FL due to prior agreements within research projects. These participants are considered trustworthy and contribute to the entire FL training process with complete information. The second approach relies on incentive-driven participation, where FL proposers solicit collaboration through incentives~\cite{zhan2021survey, tu2022incentive} from a decentralized community. In this case, participation is voluntary, and the trustworthiness of potential candidates is initially unknown.

Due to diverse availability of local resources and interests, FL participants may have different motivation in joining an FL application, and deliver diverse quality in contributions. This heterogeneity introduces several challenges, including unequal participation~\citep{xu2024reciprocal}, free-riding~\citep{fraboni2021free}, and difficulties in accurately assessing and rewarding individual contributions~\citep{witt2022decentral}. Encouraging active engagement from diverse stakeholders, such as data owners, infrastructure providers, and model developers, remains a critical challenge. This requires incentivizing participation in customized FL workflows, facilitating local model update sharing, and ensuring fair contributions to global model development within a decentralized community.

Furthermore, these complexities create efficiency challenges, as substantial time and effort are needed to set up FL workflows, identify suitable participants, and maintain learning quality. In addition, FL operations are vulnerable to both intentional and unintentional threats, including data poisoning, malicious servers, inference attacks, system disruptions, and service unavailability~\cite{mothukuri2021survey}. Addressing these risks is particularly difficult when some participants are unreliable during the training process.

In this work, we address the efficiency challenges in FL development and management within decentralized community settings. We begin by automating FL operations by modeling them as workflows and leveraging workflow engines to orchestrate their execution on remote infrastructures at scale. Our approach emphasizes workflow-based FL management, focusing on FL workflow composition, data owner incentivization, and reliable training processes via combining with decentralized technologies.

The reminder of this article is structured as follows: Section~\ref{sec:reliableFL-related work} reviews the state-of-the-art reliable FL practices, compares with existing work, and analyzes their research gap. 
Section~\ref{sec: reliableFL-workflow} describes how we can describe FL as a computational workflow and scale it out to decentralized infrastructure via the CWL open standards. 
Section~\ref{sec:reliableFL-problem} introduces a decentralized collaboration framework for FL and illustrates the mechanism design approach and studies the contract optimality to answer research questions. 
Section~\ref{sec:reliableFL-evaluation} details the experiments and the analysis of results; finally, Section~\ref{sec:reliableFL-conclusion} concludes this work.

\section{Related Work}\label{sec:reliableFL-related work}

This section reviews the state-of-the-art reliable federated learning operation practices and explores existing frameworks for managing FL in decentralized infrastructures. 



In recent years, several FL frameworks have made notable advancements in managing the FL research and development lifecycle. \citet{yang2023flscalize} introduced FLScalize, which extends the machine learning operations (MLOps) concept to generate a baseline model, integrating FL clients, the FL server, and model performance to oversee the entire FL lifecycle. \citet{colonnelli2022federated} employed the open standard workflow language, CWL~\cite{amstutz2016cwl}, to abstract and automate FL applications in a hybrid Cloud-HPC environment. Despite these advances, their approach remains constrained by the traditional client-server architecture of centralized FL, although the open standard workflow language holds significant promise for describing diverse FL workflows~\cite{kontomaris2023cwl}.

\citet{daga2023flame} introduced the Flame framework, offering flexibility in the topology configuration of FL applications tailored to the specific deployment context using new high-level abstraction topology graphs (TAGs) that incorporate five types of FL topologies. While both \citet{colonnelli2022federated} and \citet{daga2023flame} align closely with our study by utilizing high-level abstractions to enable flexibility in topology-aware FL, a key assumption in many existing works is that potential participants will join the FL process unconditionally and are sufficiently trustworthy to contribute throughout the entire FL training process. Our study diverges from previous works in several critical ways. We explore classic workflow patterns~\citep{van2003workflow} and FL architectural patterns~\citep{lo2022architectural}, making fundamental workflow concepts --- such as automation, scalability, abstraction, portability, flexibility, and reusability --- applicable to the context of federated learning operations (FLOps) via the open standard CWL. Preliminary results of this approach have been presented in CWL-FLOps~\citep{kontomaris2023cwl}.

Furthermore, \citet{cheng2022federated} introduced a novel methodology called FLOps for managing the FL lifecycle continuously and efficiently. FLOps integrates a range of processes, technologies, and tools to enhance the efficiency and quality of developing and deploying cross-silo FL systems. They highlight that workflow-related approaches, such as metadata engineering, dual deployment, and checkpoints, can help establish and automate FLOps practices, enabling smoother and more efficient operations from an engineering perspective. However, the automation of FLOps remains an open challenge, particularly when addressing security and privacy concerns, which differ from those encountered in traditional DevOps~\citep{ebert2016devops} and MLOps~\citep{kreuzberger2023machine} practices. Moreover, they do not address the issue of incentivizing participants to make high-quality contributions if the assumption of their willingness and trustworthiness does not hold --- this is a crucial factor for maintaining a reliable and secure FL process.

Our work connects to the broader discussion of promoting collaborative fairness among federated participants within decentralized communities, where incentives and fairness have been extensively explored~\citep{zeng2021comprehensive, witt2022decentral, haghtalab2025platforms}. \citet{wang2023incentive} examined incentive mechanism design in the context of resource allocation for FL clients in Blockchain-based Federated Learning (BCFL). They modeled the problem as a two-stage Stackelberg game under both complete and incomplete information. Using the Shapley value approach, they quantified clients’ contributions to the training process. By transforming the game model into two optimization problems and solving them sequentially, they derived the optimal strategies for both players. \citet{gao2022fgfl} introduced FGFL, an attack-resistant incentive mechanism for FL that detects and repels abnormal updates, thereby protecting the system in unreliable scenarios. The task publisher rewards efficient workers and punishes or eliminates malicious ones based on reputation and contribution indicators.

While both incentive mechanisms are deemed feasible through experimental evaluation, neither explores the optimal contract design required for creating smart contracts for FL operations. The question of whether the incentives embedded in smart contracts are either over-rewarding or insufficient to effectively motivate participants remains unclear. \citet{kang2019incentive} proposed a joint optimization approach combining a reputation-based worker selection scheme with contract theory to determine the optimal computation resources (e.g., contributed CPU cycles) and corresponding rewards for all participants. However, this approach does not fully capture the entire contribution to the FL training process. Furthermore, although existing works touch on fair rewards, there is a clear need for precise fairness metrics that align with specific incentives, particularly in the context of reliable FL systems.

\section{Federated Learning as a Workflow}\label{sec: reliableFL-workflow}
This section explores the advantages of managing federated learning as a workflow within decentralized infrastructure, addressing the following research question: "How can we abstract FL workflows in an open standard manner and organize their execution on remote infrastructure?"

\subsection{Federated Learning}

We consider an FL application scenario consisting of $N=\{1,2,\cdots, n\}$ clients, each client $i$ holds a local dataset $D_i = \{(x,y)\}\thicksim \mathcal{D}_{i}$ consisting of $s_i$ data samples, where $x$ and $y$ denote the data sample and its corresponding labels, respectively~\cite{zhou2024understanding}. The total sample size $S_{\text{sample}}$ from $N$ clients is given by $S_{\text{sample}} = \sum_{i=1}^{n}s_i, \forall i\in N$. Additionally, the union of all client datasets denote by $D = \cup_{i=1}^n D_i \thicksim \mathcal{D}$. FL aims to minimize a global loss function $\mathcal{L}(\cdot)$ from the $N$ clients' local loss function $\mathcal{L}_{i\in N}(\cdot)$ through the model aggregation strategy techniques. The overall FL problem can be formulated as 
\begin{align}\label{eq:FLloss}
    \min_{\mathbf{w} \in \mathbb{R}} \mathcal{L}(\mathbf{w}) \quad \text{ where } \sum_{i=1}^N \frac{s_i}{S_{\text{sample}}} \mathcal{L}_i(\mathbf{w})
\end{align}
Here, $\mathcal{L}_i(\mathbf{w}) = \sum_{k=1}^{s_i}\ell_i(\mathbf{w};(x_k, y_k)\in D_i) $ is the expected local loss of the $i^{th}$ client on its local dataset $D_i$, and $\ell_i(\mathbf{w}; (x_k, y_k))$ is the local loss function of the shared model weights $\mathbf{w}$ on sample data $(x_k, y_k)\in D_i$.

Traditionally, clients train their local models $\mathbf{w}^{(t)}_i$ independently at round $t$ by optimizing the loss function $\mathcal{L}_i(\cdot)$ on their local datasets. Clients validate the trained model using a validation dataset and upload their updated models to an aggregator for federated model aggregation. The aggregator then performs the aggregation strategy and updates the shared global model of all the local model updates. Take the model averaging method, i.e., FedAvg~\cite{mcmahan2017communication}, as an example. It is a technique developed to reduce the variance of a global model update by periodically averaging models trained over multiple communication rounds in FL. The model averaging aggregation is often formulated as,
\begin{equation}\label{eq:FMA}
    \mathbf{w}^{(t+1)} = \sum_{i=1}^n\frac{s_i}{S_{\text{sample}}}\mathbf{w}^{(t)}_i
\end{equation}
where $\mathbf{w}^{(t+1)}$ denotes the updated global model weight for the new round $t+1$. Then, it will be sent back to all active clients to initialize the next round of local training, iteratively. This process repeats until the global loss converges. 

It is known that high-quality data, efficient computation, and reliable communication at each round may contribute to the local model training with high model performance (e.g., high accuracy) and can lead to faster convergence of the local loss function $\mathcal{L}_i(\mathbf{w})$. The faster speed of the convergence of the local training and high-quality model updates will make the convergence of the global loss function quicker, and thus, the training time and cost will decrease for a targeted model performance~\cite{kang2019incentive}. Therefore, participants with high-quality data, high-performance computation, and communication efficiency can significantly improve the FL quality and efficiency. 


\subsection{Federated Learning as a Workflow}
We define an FL workflow $F = (\mathbf{w}, N, A, tT)$ as a combination of four random sets of variables: the global model $\mathbf{w}$, the participant set $N$, the aggregation strategy $A$, and the topology type $tT$. These variables consist of the fundamental building blocks of FL workflows, and users can reconfigure such blocks and customize $F$ with their preferences under an agreement of all participants. 

We build on existing solutions to develop a flexible, adaptable approach tailored to end users' specific needs. 
As a first step, we explored the component containerizer in NaaVRE~\cite{zhao2022notebook} to create and store FL building blocks --- such as models and aggregation strategies --- as research assets. For instance, a model developer can build ML models in a local NaaVRE environment using small-scale datasets, performing testing and validation. The model provider can then package the ML model or code files with the FL framework into a container and store them on Docker Hub. It facilitates the global distribution of reusable, containerized applications~\cite{launet2023federating}. This approach enhances reproducibility, portability, and scalability, streamlining the integration of these building blocks into reconfigurable FL workflows~\cite{wang2022scaling} and accelerating FL development.

In~\cite{launet2023federating}, we primarily used docker-nvidia for local client training with CUDA GPU resources, automating FL pipelines while allowing clients to retain control over their data and computation. For large-scale automated deployment, Docker Swarm serves as an alternative solution.
The case study on histological image analysis demonstrated the feasibility of the proposed approach, where we utilized the Flower framework~\cite{beutel2020flower} within the Jupyter environment for FL code development. However, most existing frameworks, such as TensorFlow Federated (TFF)\footnote{\url{https://www.tensorflow.org/federated}}, NVFlare\footnote{\url{https://github.com/NVIDIA/NVFlare}}, IBM FL~\cite{ludwig2020ibm}, OpenFL~\cite{reina2021openfl}, PySyft~\cite{ziller2021pysyft}, and Flower, follow a centralized FL paradigm with a client-server architecture, in which the aggregator cannot be replaced or changed during the training process. They typically rely on direct bidirectional communication (e.g., gRPC~\cite{indrasiri2020grpc}) between the aggregator and client training nodes, which limits their flexibility in supporting diverse FL deployment scenarios. For example, locations of decentralized data infrastructure can be heterogeneous, exposing different hardware resources and protocols for authentication, communication, resource allocation, and job execution. Plus, they can be independent of each other, meaning that direct communication among them may not be allowed~\cite{colonnelli2022federated}.

To overcome this limitation, we model the FL pipeline as a computational workflow and utilize a workflow engine to orchestrate its execution across remote infrastructures. Specifically, we employ CWL open standards to define FL pipelines, and make them more manageable to automate and parallelize the joint model training from the local computer to the remote cloud environments. 
We first employ the classic workflow patterns to describe diverse computational FL workflows. Table~\ref{tab:mapping} introduces a taxonomy of 43 control patterns and 40 data patterns identified in the Workflow Patterns Initiative (WPI)\footnote{\url{http://www.workflowpatterns.com/}}, mapping them to a diverse range of FL topology types.


\renewcommand\theadalign{bc}
\renewcommand\theadfont{\bfseries}
\renewcommand\theadgape{\Gape[4pt]}
\renewcommand\cellgape{\Gape[4pt]}

\begin{table*}[!htb]
\scriptsize
    \centering
    \setlength\tabcolsep{2pt}
    \caption{The mappings of different FL topology types on WPI's workflow patterns. 
    }\label{tab:mapping}
    {\renewcommand{\arraystretch}{1.25}%
    \resizebox{\textwidth}{!}{

    \begin{tabular}{|ccc|cccccccccccc|}
\hline
\multicolumn{3}{|c|}{\multirow{3}{*}{}}                                                                                                                & \multicolumn{12}{c|}{FL   Topology Types}                                                                                                                                                                                                                                                                                                                                \\ \cline{4-15} 
\multicolumn{3}{|c|}{}                                                                                                                                 & \multicolumn{2}{c|}{Star}                                            & \multicolumn{2}{c|}{Tree}                                        & \multicolumn{3}{c|}{Decentralized}                                                                & \multicolumn{5}{c|}{Minor}                                                                                                 \\ \cline{4-15} 
\multicolumn{3}{|c|}{}                                                                                                                                 & \multicolumn{1}{c|}{\rotatebox{60}{Synchronous}} & \multicolumn{1}{c|}{\rotatebox{60}{Asynchronous}} & \multicolumn{1}{c|}{\rotatebox{60}{Hierarchical}} & \multicolumn{1}{c|}{\rotatebox{60}{Dynamic}} & \multicolumn{1}{c|}{\rotatebox{60}{De-Mesh}} & \multicolumn{1}{c|}{\rotatebox{60}{De-Wireless}} & \multicolumn{1}{c|}{\rotatebox{60}{Blockchain}} & \multicolumn{1}{c|}{\rotatebox{60}{Ring}} & \multicolumn{1}{c|}{\rotatebox{60}{Clique}} & \multicolumn{1}{c|}{\rotatebox{60}{Grid}} & \multicolumn{1}{c|}{\rotatebox{60}{Fog}} & \rotatebox{60}{Semi-Ring} \\ \hline
\multicolumn{1}{|c|}{\multirow{13}{*}{\rotatebox{90}{Patterns}}} & \multicolumn{1}{c|}{\multirow{8}{*}{\rotatebox{90}{Control (43)}}} & Basic control (5/5)                            & \multicolumn{1}{c|}{1}           & \multicolumn{1}{c|}{1}            & \multicolumn{1}{c|}{1}            & \multicolumn{1}{c|}{1}       & \multicolumn{1}{c|}{1}       & \multicolumn{1}{c|}{1}           & \multicolumn{1}{c|}{1}          & \multicolumn{1}{c|}{1}    & \multicolumn{1}{c|}{1}      & \multicolumn{1}{c|}{1}    & \multicolumn{1}{c|}{1}   & 1         \\ \cline{3-15} 
\multicolumn{1}{|c|}{}                           & \multicolumn{1}{c|}{}                              & Advanced Branching and Synchronization (1/14)  & \multicolumn{1}{c|}{1}           & \multicolumn{1}{c|}{1}            & \multicolumn{1}{c|}{1}            & \multicolumn{1}{c|}{1}       & \multicolumn{1}{c|}{1}       & \multicolumn{1}{c|}{1}           & \multicolumn{1}{c|}{1}          & \multicolumn{1}{c|}{1}    & \multicolumn{1}{c|}{1}      & \multicolumn{1}{c|}{1}    & \multicolumn{1}{c|}{1}   & 1         \\ \cline{3-15} 
\multicolumn{1}{|c|}{}                           & \multicolumn{1}{c|}{}                              & Iteration (3/3) --\textgreater loops extension & \multicolumn{1}{c|}{1}           & \multicolumn{1}{c|}{1}            & \multicolumn{1}{c|}{1}            & \multicolumn{1}{c|}{1}       & \multicolumn{1}{c|}{1}       & \multicolumn{1}{c|}{1}           & \multicolumn{1}{c|}{1}          & \multicolumn{1}{c|}{1}    & \multicolumn{1}{c|}{1}      & \multicolumn{1}{c|}{1}    & \multicolumn{1}{c|}{1}   & 1         \\ \cline{3-15} 
\multicolumn{1}{|c|}{}                           & \multicolumn{1}{c|}{}                              & Multiple Instance (1/7)                        & \multicolumn{1}{c|}{1}           & \multicolumn{1}{c|}{1}            & \multicolumn{1}{c|}{1}            & \multicolumn{1}{c|}{1}       & \multicolumn{1}{c|}{0}       & \multicolumn{1}{c|}{1}           & \multicolumn{1}{c|}{1}          & \multicolumn{1}{c|}{0}    & \multicolumn{1}{c|}{1}      & \multicolumn{1}{c|}{1}    & \multicolumn{1}{c|}{1}   & 1         \\ \cline{3-15} 
\multicolumn{1}{|c|}{}                           & \multicolumn{1}{c|}{}                              & State-based (0/5)                              & \multicolumn{1}{c|}{0}           & \multicolumn{1}{c|}{1}            & \multicolumn{1}{c|}{0}            & \multicolumn{1}{c|}{1}       & \multicolumn{1}{c|}{1}       & \multicolumn{1}{c|}{1}           & \multicolumn{1}{c|}{1}          & \multicolumn{1}{c|}{1}    & \multicolumn{1}{c|}{1}      & \multicolumn{1}{c|}{1}    & \multicolumn{1}{c|}{1}   & 1         \\ \cline{3-15} 
\multicolumn{1}{|c|}{}                           & \multicolumn{1}{c|}{}                              & Trigger (0/2)                                  & \multicolumn{1}{c|}{1}           & \multicolumn{1}{c|}{1}            & \multicolumn{1}{c|}{1}            & \multicolumn{1}{c|}{1}       & \multicolumn{1}{c|}{1}       & \multicolumn{1}{c|}{1}           & \multicolumn{1}{c|}{1}          & \multicolumn{1}{c|}{1}    & \multicolumn{1}{c|}{1}      & \multicolumn{1}{c|}{1}    & \multicolumn{1}{c|}{1}   & 1         \\ \cline{3-15} 
\multicolumn{1}{|c|}{}                           & \multicolumn{1}{c|}{}                              & Cancelation and Force Completion (0/5)         & \multicolumn{1}{c|}{0}           & \multicolumn{1}{c|}{1}            & \multicolumn{1}{c|}{1}            & \multicolumn{1}{c|}{1}       & \multicolumn{1}{c|}{1}       & \multicolumn{1}{c|}{1}           & \multicolumn{1}{c|}{0}          & \multicolumn{1}{c|}{0}    & \multicolumn{1}{c|}{0}      & \multicolumn{1}{c|}{1}    & \multicolumn{1}{c|}{1}   & 1         \\ \cline{3-15} 
\multicolumn{1}{|c|}{}                           & \multicolumn{1}{c|}{}                              & Termination (1/2)                              & \multicolumn{1}{c|}{0}           & \multicolumn{1}{c|}{0}            & \multicolumn{1}{c|}{1}            & \multicolumn{1}{c|}{1}       & \multicolumn{1}{c|}{1}       & \multicolumn{1}{c|}{1}           & \multicolumn{1}{c|}{0}          & \multicolumn{1}{c|}{0}    & \multicolumn{1}{c|}{1}      & \multicolumn{1}{c|}{1}    & \multicolumn{1}{c|}{1}   & 1         \\ \cline{2-15} 
\multicolumn{1}{|c|}{}                           & \multicolumn{1}{c|}{\multirow{5}{*}{\rotatebox{90}{Data (40)}}}    & Data Visibility (4/8)                          & \multicolumn{1}{c|}{1}           & \multicolumn{1}{c|}{1}            & \multicolumn{1}{c|}{1}            & \multicolumn{1}{c|}{1}       & \multicolumn{1}{c|}{0}       & \multicolumn{1}{c|}{1}           & \multicolumn{1}{c|}{1}          & \multicolumn{1}{c|}{1}    & \multicolumn{1}{c|}{1}      & \multicolumn{1}{c|}{1}    & \multicolumn{1}{c|}{1}   & 1         \\ \cline{3-15} 
\multicolumn{1}{|c|}{}                           & \multicolumn{1}{c|}{}                              & Internal Data Interaction (5/6)                & \multicolumn{1}{c|}{1}           & \multicolumn{1}{c|}{1}            & \multicolumn{1}{c|}{1}            & \multicolumn{1}{c|}{1}       & \multicolumn{1}{c|}{0}       & \multicolumn{1}{c|}{1}           & \multicolumn{1}{c|}{1}          & \multicolumn{1}{c|}{1}    & \multicolumn{1}{c|}{1}      & \multicolumn{1}{c|}{1}    & \multicolumn{1}{c|}{1}   & 1         \\ \cline{3-15} 
\multicolumn{1}{|c|}{}                           & \multicolumn{1}{c|}{}                              & External Data Interaction (0/12)               & \multicolumn{1}{c|}{1}           & \multicolumn{1}{c|}{1}            & \multicolumn{1}{c|}{1}            & \multicolumn{1}{c|}{1}       & \multicolumn{1}{c|}{0}       & \multicolumn{1}{c|}{1}           & \multicolumn{1}{c|}{1}          & \multicolumn{1}{c|}{1}    & \multicolumn{1}{c|}{1}      & \multicolumn{1}{c|}{1}    & \multicolumn{1}{c|}{1}   & 1         \\ \cline{3-15} 
\multicolumn{1}{|c|}{}                           & \multicolumn{1}{c|}{}                              & Data Transfer (3/7)                            & \multicolumn{1}{c|}{1}           & \multicolumn{1}{c|}{1}            & \multicolumn{1}{c|}{1}            & \multicolumn{1}{c|}{1}       & \multicolumn{1}{c|}{0}       & \multicolumn{1}{c|}{1}           & \multicolumn{1}{c|}{1}          & \multicolumn{1}{c|}{1}    & \multicolumn{1}{c|}{1}      & \multicolumn{1}{c|}{1}    & \multicolumn{1}{c|}{1}   & 1         \\ \cline{3-15} 
\multicolumn{1}{|c|}{}                           & \multicolumn{1}{c|}{}                              & Data-based Routing (2/7)                       & \multicolumn{1}{c|}{1}           & \multicolumn{1}{c|}{1}            & \multicolumn{1}{c|}{1}            & \multicolumn{1}{c|}{1}       & \multicolumn{1}{c|}{1}       & \multicolumn{1}{c|}{1}           & \multicolumn{1}{c|}{1}          & \multicolumn{1}{c|}{1}    & \multicolumn{1}{c|}{1}      & \multicolumn{1}{c|}{1}    & \multicolumn{1}{c|}{1}   & 1         \\ \hline
\end{tabular}
    }

    }
\end{table*} 

A review of the 43 WPI patterns related to the control-flow perspective reveals that 11 patterns are supported by CWL v1.2 or earlier. Of the 40 data patterns, only 17 are supported by CWL constructs, and just 2 of the 43 resource patterns are covered. Additionally, CWL standards offer limited support for exception handling and do not address the event log imperfection patterns defined in the WPI patterns. For workflow presentation patterns, 19 are supported by CWL v1.2 or earlier, with one pattern still unsupported. Certain environments, such as blockchain networks, mobile devices, and wireless systems, cannot directly interface with CWL-supported workflow patterns. 

\begin{figure}[!htb]
    \centering
    \includegraphics[width=.96\linewidth]{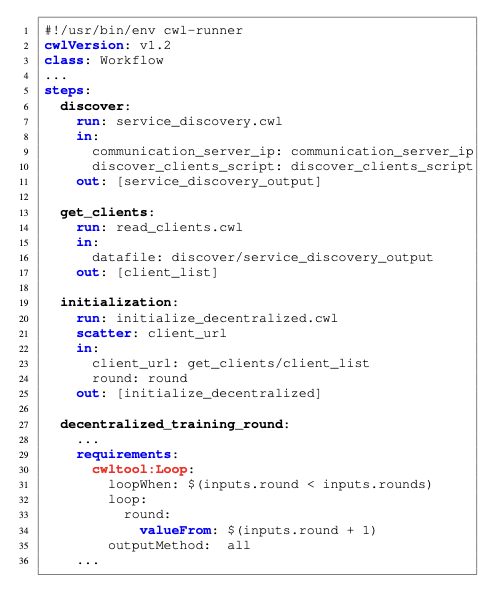}
    \caption{Snippet of the CWL workflow.}
    \label{fig:snippet-fl}
\end{figure}

FL requires loop patterns for describing the iterative training. While CWL v1.2 lacks loop constructs and recursion support, there are alternative methods to enable loops in describing iterative FL workflows. These include 1) combining sub-workflows with loop extensions in the current stable version (v1.2) and 2) the upcoming v1.3.0-dev1 version of CWL, which is expected to introduce loop constructs in a future stable release. Figure~\ref{fig:snippet-fl} shows the snippet of the main steps of a decentralized FL workflow example written in CWL version v1.2. It orchestrates multiple steps, including service discovery, client reading, initialization, and decentralized training rounds. This CWL snippet set up a loop within a sub-workflow named \texttt{decentralized\_training\_round}. It manages decentralized training rounds in FL, iterating until the specified number of rounds is completed. Unlike a fixed aggregator in the centralized FL workflow, the aggregator is randomly chosen from the distributed clients to perform federated model aggregation in the decentralized training round design. 
The loop feature implements a dynamic mechanism to handle multiple training rounds, ensuring scalability and flexibility in FL operations. Additionally, it can be fully automated using GitHub Action workflows. More details can be found in the source code, which is available in the GitHub repository\footnote{\url{https://github.com/CWL-FLOps/DecentralizedFL-CWL/}}.



\subsection{Demonstration of FL Workflow}
We demonstrate how the workflow-based approach enables scalable automation of FL operations across cloud infrastructures. Additionally, we integrate the implemented CWL-supported FL workflows with the Jupyter environment to facilitate collaborative learning and seamless workflow management. Some related preliminary results have been published in~\cite{krishnasamy2024collaborative, kontomaris2023cwl}. 

\begin{figure}[!htb]
    \centering
    \includegraphics[width=.96\linewidth]{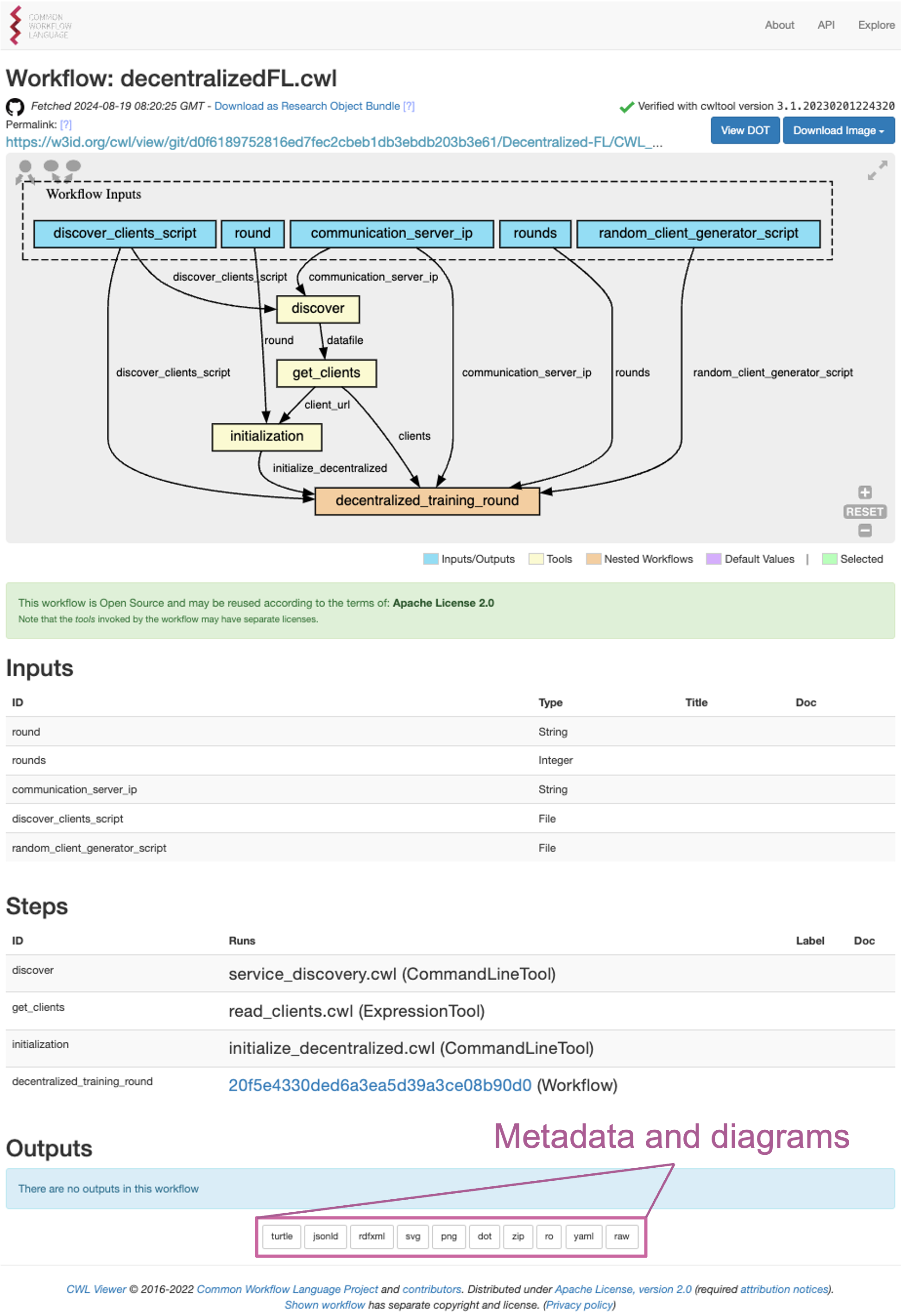}
    \caption{The screenshot of the visualized FL workflow in the CWL viewer with detailed license, specification, and metadata.}
    \label{fig:CWL_FL}
\end{figure}

\vskip 0.1cm
\noindent \textbf{Visualized CWL-supported FL workflow.}
Figure~\ref{fig:CWL_FL} displays the screenshot of the visualized CWL workflow graph named \texttt{decentralizedFL.cwl} via the CWL viewer tool\footnote{\url{https://view.commonwl.org/}}. 
It has been verified with cwltool version 3.1.20230201224320 and consists of five inputs (highlighted in a blue rectangle), three steps (highlighted in a yellow rectangle), and a nested workflow (highlighted in an orange rectangle). The connections between the workflow inputs and steps illustrate the dependencies within the process. This workflow can serve as a research object bundle, enriched with comprehensive metadata and licensing information to ensure its reusability. For example, it is publicly available as an open-source workflow and can be reused under the terms of the Apache License 2.0.

\vskip 0.1cm
\noindent \textbf{FL workflow execution on hybrid Clouds.}
To demonstrate the FL workflow execution over decentralized data infrastructures, we utilized a well-known dataset named MNIST\footnote{\url{https://git-disl.github.io/GTDLBench/datasets/mnist_datasets/}} and five cloud instances from different providers. These included one \textit{t4g.small} instance (2 ARM-based vCPUs, 2 GiB memory) and \textit{t3.small} instance (2 x86-based vCPUs, 2 GiB memory) from Amazon Web Services (AWS), one \textit{t2a-standard-1} instance (1 ARM-based vCPU, 4 GiB memory) and \textit{e2-medium} instance (2 x86-based vCPUs, 4 GiB memory) from Google Cloud Platform (GCP), and one \textit{lab.uvalight.net} instance (2 x86-based vCPUs, 2 GiB memory) from the OpenLab infrastructure provided by the university. Each dataset stored on the cloud instances was split from the MNIST dataset, with a size of approximately 60 MB. The FL source code is based on PyTorch, reproduced from~\cite{shaoxiong_ji_2018_4321561}, where the author used the FedAvg method for the federated model aggregation. We re-configured the FL workflow with ten local training epochs and 12 communication rounds. Each setup was executed ten times to collect results for statistical analysis. 

 
The average training time is approximately $(9.786 \pm 0.238)$ minutes, and the average test accuracy achieved by the FL workflow execution is $(98.127 \pm 0.085)\%$. 
These empirical results demonstrate that CWL-supported FL workflows are scalable, reusable, and portable, making them suitable for execution in off-chain decentralized infrastructures such as hybrid cloud environments. Since the CWL community provides standardized alternatives for defining portable and reusable workflows that remain engine- and vendor-neutral, any CWL-compliant workflow execution engine should be able to execute this standardized workflow description and produce consistent results, regardless of the underlying infrastructure~\cite{goble2020fair}. Furthermore, by incorporating the principles of FAIR computational workflows within CWL, we can advance toward realizing FAIR FL workflow management in future work.

\subsection{Lessons and Challenges}\label{sec: reliableFL-le&ch}
The empirical results demonstrate that CWL effectively describes FL workflows and automates their execution using cloud technologies. However, this approach assumes a pre-established collaboration, where distributed data sources are predefined within the workflow. 

CWL does not inherently address trust and reliable collaboration among multiple participants in FL workflows. The decentralized nature of data providers challenges the traditional centralized workflow management paradigm, where providers are predefined for composing application-specific FL workflows. To overcome this, an effective collaboration framework is needed to enable the dynamic selection of data or resource providers and ensure the quality of their contributions.

In the next section, we will explore how to incentivize collaborations and ensure training quality by addressing the following research questions:

\begin{itemize} 
\item How can we incentivize participants with high-quality contribution to join the FL training process? 
\item How can we assure training quality from decentralized participants in an unreliable environment? 
\end{itemize}

\section{Decentralized Collaboration Framework}\label{sec:reliableFL-problem}
To tackle the challenges identified in~\ref{sec: reliableFL-le&ch}, this section introduces a decentralized collaboration framework that aims to manage the dynamic collaboration among participants within a decentralized community. The basic design ideas include:
\begin{itemize}
    \item Using blockchain environment to manage the decentralized relation among FL participants and to record their interaction history.
    \item Managing the collaboration agreement by employing the smart contract of the blockchain, where an FL developer can explicitly describe the conditions for desired participants in a smart contract.
    \item Verifying the contribution through blockchain ledgers and manage the update of the FL aggregations via dynamic consensus among participants through smart contracts. 
\end{itemize}


\subsection{System Overview}
\begin{figure*}[!htb]
    \centering
    \includegraphics[width=.96\linewidth]{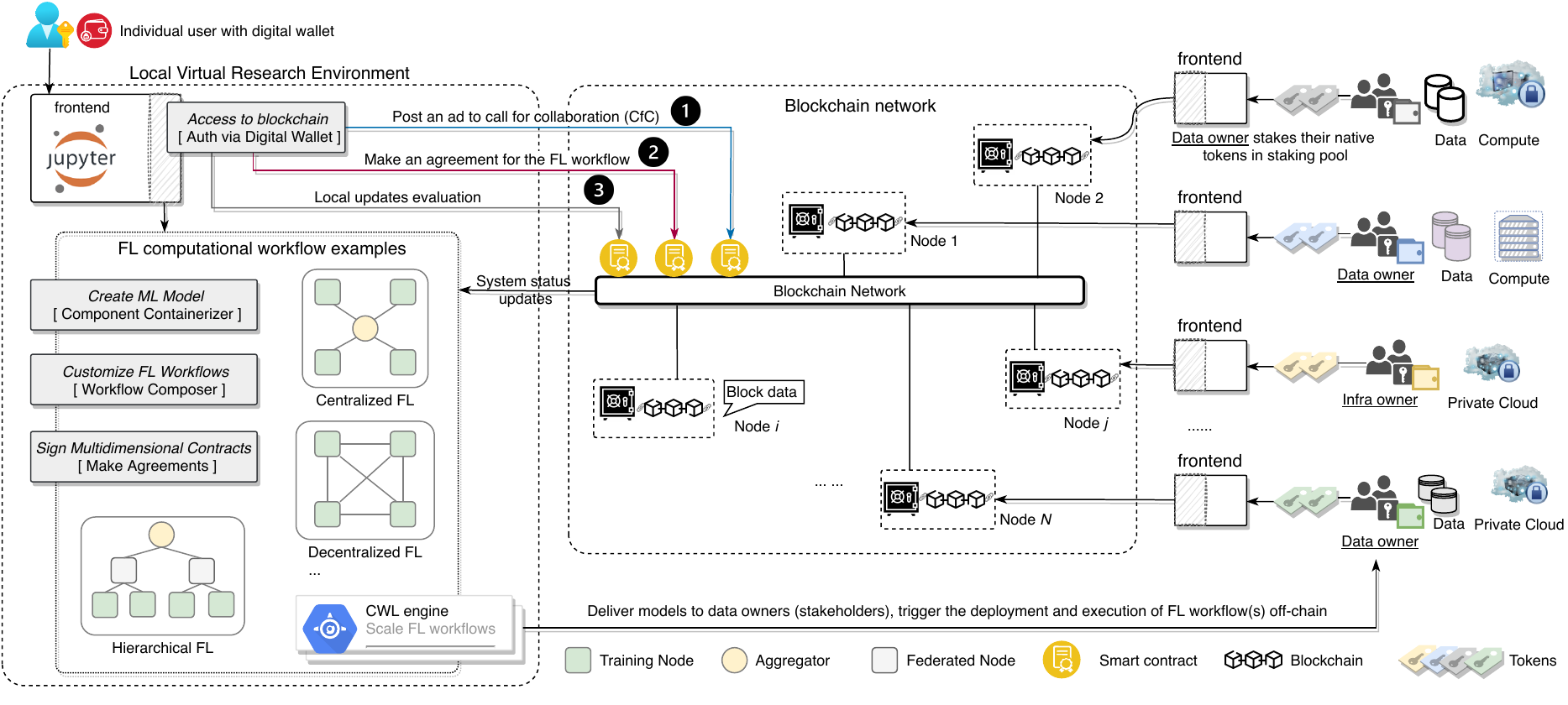}
    \caption{This is an overview of the system for FL applications. A local virtual research environment enables a user to develop ML models on-premise and unlocks the potential of establishing robust collaboration in a decentralized environment via~\ding{182}~\ding{183} and~\ding{184}. It consists of two main interfaces: 1) via blockchain interfaces, users can access to the blockchain network to call for collaboration, make an agreement, and evaluate local updates on-chain; 2) via workflow runtime interfaces, users can create ML models, customize FL workflows, operate collaborative workflows on decentralized infrastructure.}
    \label{fig:sys-model}
\end{figure*}

Figure~\ref{fig:sys-model} depicts the basic idea for FL applications, building on our previous work on the D-VRE framework~\cite{wang2024decentralized}. 

This framework equips Jupyter users with essential components for enabling robust FL collaboration, integrating a personal Jupyter working environment, a blockchain network, and off-chain legacy resources, such as data storage, computation, and networking, within a decentralized ecosystem. 
It enables a group of Jupyter users with different roles to engage in scientific community activities and create FL building blocks as research assets. This is achieved through the \textit{Create FL Building Blocks} function (Component Containerizer within NaaVRE).

Through the function \textit{Access to blockchain} (Auth via Digital Wallet within D-VRE), the system allows users to configure their digital wallets like MetaMask, to unlock the decentralized Web and applications to call for collaboration and secure asset sharing in a decentralized environment~\cite{wang2024decentralized}. 

Due to the diverse data access restrictions, data can not be moved out of the institutions~\cite{launet2023federating}. The system allows model providers (also called FL task publishers) to Call for Collaborations (CfCs) to enable secure data sharing, computing resource sharing, and collaborative learning from the decentralized community, via the \textit{Sign multidimensional contracts} function (Make Agreements within D-VRE). Data or resource owners can respond to the CfCs via the interfaces embedded in their personal frontends. 

Each CfC will include specific requirements, e.g., the data and resource specifications, required stakes, targeted model performance, minimum number of participants, and what the publisher can offer (such as reward), rules of penalty, and reputation updates resulting from different contribution behaviours. These requirements can be explicitly described in a smart contract. The stake is related to crypto staking in the blockchain system, which is the practice of locking participants' digital tokens to a blockchain network to earn rewards. For instance, a participant's stake can be returned later with an additional reward if there is no indication of malicious behaviours~\cite{nguyen2024stake}. The stake plays a crucial role in reliable FL, as 1) it can increase the cost of attacks and 2) promote risk-sharing during FL system operations. 
The candidate has to join the contract to make agreements. 


Via the function of \textit{Customize FL workflow} (Workflow Composer and CWL-based FL workflow composition), users who signed contracts acting as FL participants can customize a computational FL workflow through verified resources from the off-chain decentralized (data) infrastructures. We assume that data are stored in the infrastructures which provide computation capacity and network for communication. Some participants may only contribute to infrastructure, e.g., computing, storage, and network resources. Each participant can be identified and verified for their resources through trusted external committees, such as Oracle and DAO~\cite{wang2019decentralizedautonomousorganizations}. Hence, the composition of computational FL workflows can combine the on-chain and off-chain services to scale different FL scenarios. 


After the FL workflow with on-chain engagement has been confirmed, the system generates a number of smart contract instances to enforce the contract items and ensures that precise, mutually agreed-upon terms govern collaborations in the decentralized network. The system enables users to start the CWL engine for automating FL workflow deployment and execution over legacy data infrastructures while enabling secure and transparent execution in the blockchain environment. Since the block data has stored ledgers about FL actions for all participants in the blockchain network, it can correlate the captured data with user behaviours to maintain a transparent record of any changes made to an FL operation. Finally, users can evaluate local updates of their FL workflows related to domain-specific subjects and analyze the consequences of participants' actions regarding relevant task contributions, reputation updates, and rewards.

\subsection{How to Incentivize Collaboration?}
In real-life scenarios, the FL task publisher does not know which users in the system would join the FL training due to the lack of prior information. The local data quality and the computing power of available resources from potential candidates are unknown to the publisher owing to data privacy and security concerns. Therefore, it is essential for the task publisher to design an efficient incentive mechanism to stimulate active engagement from high-quality candidates while reducing the over-reward risks caused by the issues of asymmetric information~\cite{kang2019incentive, ding2020optimal}. 
This study employs the contract theory with a multidimensional scheme as an efficient approach to address the incentive collaboration problem.

Assume a set of $\mathbb{N}$ potential candidates responds to the CfC task in the system. Let $N = {1, 2, \cdots, n}$, where $N \subseteq \mathbb{N}$, represent the set of identity-verified candidates who have signed contracts with stakes $S = {S_1, S_2, \cdots, S_n}$ and initial reputations ${r^0} = {r^0_1, r^0_2, \cdots, r^0_n}$. These candidates act as FL participants, continuously contributing to the FL network. Each participant's reputation is dynamically updated based on their behaviour.

Let $\theta$ denote a standard type space to differentiate the participant nodes. Suppose participants are classified into $N$ types, which can be sorted in a descending order regarding contribution types: $\theta_1>\theta_2>\cdots >\theta_n, i\in\{1,2, \cdots, n\}$. Let $\widehat{\theta}_i$ be the claimed type by participant $i$ with the true $\theta_i$. We assume that candidates who join the FL must claim their types $\widehat{\theta}$ and select the corresponding contract item $\phi_{\widehat{\theta}}$. Although the task publisher does not know about the true type of a given participant due to the information asymmetry, it has information about the reported type $\widehat{\theta}_i$ by the participant with the true type $\theta_i$ (e.g., data and infrastructure specifications), from which the publisher can inform the probability that a participant belongs to a particular type $\theta_i$ with reported (meta)data and computing power categories, denoted as $\sum_{i=1}^N p_i(\widehat{\theta}_i)=1$. 

For participants with different contributions, the task publisher signs different contract items with them. We define the contract $\mathcal{C}=(T_{\text{max}}, \Phi)$ is comprised of a maximum waiting time $T_{\text{max}}$ and contract terms $\Phi = \{\phi_i\}_{i\in N}$. $T_{\max}$ denotes the maximum allowable time for receiving participants' contributions. Each contract item $\phi_i \triangleq (C_i, S_i, R_i)$ specifies the relationship among each type-$i$ participant's stake $S_i$, contribution $C_i$ (often associated with a completion time $\tau_i \leq T_{\text{max}}$), and its corresponding reward $R_i$. Any participant who completes their contributions to the FL network within the required time during each communication round will receive a reward $R_i$. Conversely, participants whose completion time $\tau_i$ exceeds $T_{\text{max}}$ will receive a zero reward, resulting in a zero contract instance. 

\vskip 0.1cm 
\noindent \textbf{Contribution model.}
We define a contribution model for each FL workflow. Let $C_i^{(t)}$ denote an estimated contribution based on the $i^{th}$ participant's local model training $f_{\text{local}}$ and its behaviours $g_{\text{behaviour}}$ at round $t$, which can be formally defined as:
\begin{align}\label{eq:contrib}
    C_i^{(t)} = &f_{\text{local}}(D_i, \mathbf{w}_i^{(t)})\cdot g_{\text{behaviour}}(h_i^{(t)}), \quad \forall i \in N
\end{align}
where $C_i^{(t)}\geq 0$ and $h_i^{(t)}$ is the historical behaviour records regarding reputation and stake which can be used to differentiate honest participant nodes and malicious ones. 

Let $T_{\text{max}}$ denote the upper limit on the duration in FL synchronous settings, within which participants must submit their contributions, such as local model updates $\mathbf{w}^{(t)}i$, before the aggregation is triggered within the federation. Specifically, participants with $\tau_i \leq T_{\text{max}}$ can complete the submission of model updates to allow the aggregation of federated models to proceed on time per communication round. We define the contribution value function $V(\cdot)$ for any participant $i \in N$ as follows:
\begin{align}\label{eq:vc}
     V(C^{(t)}_i, \tau_i) =  \frac{X_c}{\tau_i} \cdot q_i^{(t)}, \quad  q_i^{(t)} = \sigma(\frac{C^{(t)}_i-C_{\min}}{C_{\max} - C_{\min}}) 
\end{align}
where $\tau_i$ is the time cost for participant $i$ to make contribution $C_i^{(t)}$, and $\frac{X_c}{\tau_i}$ represents the unit price of the contribution for participant $i$, where $X_c$ is a constant factor representing the value associated with the contribution. We define $q_i^{(t)}$ as a dynamic quality parameter via the sigma function $\sigma(\cdot)$, where $C_{\max}$ represents the saturation threshold for contribution, and $C_{\text{min}}$ is the minimum required contribution. A higher value of $V(\cdot)$ indicates better quality in the local training contribution, leading to fewer training iterations needed to reach a targeted model performance score (e.g., accuracy, precision, recall, or F1 score). 

\vskip 0.1cm
\noindent \textbf{Utility function of participants.}
For a signed contract $\phi_i$, we define the utility function of the type-$i$ participant at round $t$ as follows, 
\begin{align}
    &U_c(\phi_i|\theta_i) = R_i \mathbb{I}_{\text{compl}} - \lambda_s S_i (1-\mathbb{I}_{\text{compl}}) - c(C^{(t)}_i)
    \\
    &\mathbb{I}_{\text{compl}} = \text{exp}(-\sum_{k=1}^K \omega_k\cdot \text{VSL}_k).
\end{align}
The compliance condition, $\mathbb{I}_{\text{compl}} \in [0,1]$, is derived from the violation levels. For example, minor violations (e.g., occasional timeouts) will result in a partial retention of proceeds, while severe violations (e.g., malicious behaviour) will reduce the proceeds to zero. The normalized violation level for the $K$ violation types is denoted by $\text{VSL}_k \in [0,1]$, with $\omega_k$ representing the weight of the $k^{th}$ violation type. A value of $\mathbb{I}_{\text{compl}} = 1$ indicates no violations, while $\mathbb{I}_{\text{compl}} = 0$ signifies severe violations.

The parameter $\lambda_s$ defines the penalty factor applied to the $i^{th}$ participant's stake, $S_i$. If a node operates without violations, a proportion of the stake, $\lambda_s S_i$, is allocated to the system's risk reserve. However, if violations occur, the participant risks forfeiting the entire stake, $\lambda_s S_i$. Additionally, $c(\cdot)$ represents the effort cost for contributing $C^{(t)}_i$ with time $\tau_i$ by the type-$i$ participant. This cost can be further extended to more complex expressions by factoring in the resource utilization required for model training, validation, and aggregation.

\vskip 0.1cm
\noindent \textbf{Profit function of the task publisher.}
Task publisher's profit derived from a type-$i$ participant is influenced by multiple factors. Although a high-quality contribution can increase the task publisher's profit, it also incurs a higher reward cost for the publisher. In addition, any violations (e.g., timeouts, malicious actions, or mismatched behaviours) by participants will result in a deduction from their stake (or tokens) as penalties, which are then allocated as profits to the publisher. Therefore, we define the profit function obtained from participant $i$ as follows: 
\begin{align}
    \pi(\phi_i) = & (V(C^{(t)}_i)-R_i) \mathbb{I}_{\text{compl}} + \lambda_s S_i (1-\mathbb{I}_{\text{compl}}),
\end{align}
where $V(C^{(t)}_i)-R_i\geq 0$, the profit is positive; otherwise, the publisher risks incurring a negative profit, even if no violations occur from any participant $i$. 

To make the contract feasible, it must satisfy the following constraints simultaneously. 
\begin{definition}[Individual Rationality (IR)] \label{def:IR}
    Let $U_c(\phi_i|\theta_i)$ be the utility for the type-$i$ participant under the contract $\phi_i$. Each type-$i$ participant achieves the non-negative utility if it chooses the contract item $\phi_i$ that is designed for its own type $\theta_i$, which is given by 
    \begin{equation}\label{eq:IR}
        U_c(\phi_i|\theta_i) \geq 0.
    \end{equation}
\end{definition}

\begin{definition}[Incentive Compatibility (IC)] \label{def:IC}
    Each participant $i$ achieves the maximum utility by truthfully reporting its type $\theta_i$ if it chooses the contract item $\phi_i$ that is designed for its own type $\theta_i$ rather than other types $\theta_{-i}$ in contract items $\phi_{\theta_{-i}}$. The mechanism is incentive compatible if
    \begin{align}\label{eq:IC}
        U_c(\phi_i |\widehat{\theta}_i = \theta_i, \phi_{\theta_{-i}}) \geq U_c(\phi_i|\widehat{\theta}_i \neq \theta_i, \phi_{\theta_{-i}}),\\
        U_c(\phi_i |\widehat{\theta}_i = \theta_i, \phi_{\theta_{-i}}) \geq U_c(\phi_j|\widehat{\theta}_j = \theta_j, \phi_{\theta_{-i}}),
    \end{align}
where $\theta_{-i}\cup\theta_i = \theta$, $\theta_{-i}\neq \theta_i$, and $\theta_i\geq \theta_j$ for all participants $i,j\in N$.
\end{definition}
As a rational individual, the task publisher aims to maximize its profit in the system~\cite{xiong2020incentive}. Therefore, the total incentive problem is given by 
\begin{align}\label{eq:incentive problem}
    \max \Pi(\Phi)  =  \sum_{i=1}^{N}p_i(\widehat{\theta}_i) \cdot \pi(\phi_i),  \quad \forall \phi_i\in\Phi\\
    \text{s.t., }  \textbf{IR} (Eq.~\ref{eq:IR}) \text{ and } \textbf{IC} (Eqs.~\ref{eq:IC}, 10) \text{ constraints}. 
\end{align}
To mitigate the risks of over-rewarding or insufficient rewards, it is crucial to examine the optimality of the contract problem. 

\subsection{How to Assure Training Quality along with Contract Optimality?}

\begin{figure}
    \centering
    \includegraphics[width=.58\linewidth]{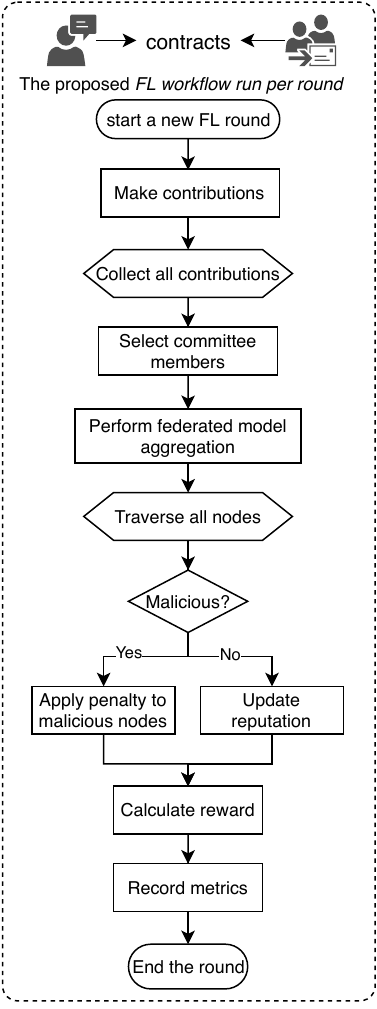}
    \caption{Flowchart of FL system operations per round.}
    \label{fig:reliableFL-flowchart}
\end{figure}

This section first employs a mechanism design approach that integrates a reputation system and dynamic incentive indicators within contracts to ensure training quality in an unreliable environment. Then, leveraging a quantified reward measurement, we analyze the optimality of the contract problem. Figure~\ref{fig:reliableFL-flowchart} illustrates the key steps involved in assuring training quality, including reputation-based committee selection, malicious detection and penalty, reputation updates, and reward calculation.

\vskip 0.1cm 
\noindent \textbf{Reputation-based committee selection.}
We introduce a committee of high-reputable and highly effective nodes among participants to do extra work like being responsible for partial model validation and aggregation. Let $\mathcal{M}^{(t)}$ be the final selected committee in each round $t$, where $\mathcal{M}^{(0)} = \emptyset$ and $\mathcal{K}$ is the committee size. We use a stratified sampling method~\cite{neyman1992two} based on reputation ${r^{(t)}}=\{r^{(t)}_1, r^{(t)}_2,\cdots, r^{(t)}_n\}$ and a cooling-down mechanism, to balance the advantages of highly reputable nodes.

First, we sort the nodes with reputation descending and divide them into several strata layers. Let $N_{\text{sorted}}=\operatorname{sort}(N, r_i\downarrow)$ be the sorted set and $L$ be the number of strata to divide the nodes into. 
For each stratum layer $L_k = N_{\text{sorted}}[\frac{(k-1)N}{L}, \frac{kN}{L}), 1\leq k\leq L$, the initial quota per stratum $Q_k$ is calculated by
\begin{equation}
    Q_k = \lfloor \frac{\mathcal{K}}{L} \rfloor+\begin{cases}
            1, \quad \quad\text{if }k \leq \mathcal{K} \mod L,\\
            0,  \quad \quad\text{otherwise}.
        \end{cases}
\end{equation}

For each layer, the set of eligible nodes is $\mathcal{E}_k=\{i\in L_k|cd^{(t)}_i = 0\}$, where $cd^{(t)}_i =0$ indicates the node $i$ is not cooling down. Hence, we can determine the number of selected nodes $m_k = \max(1, \min(Q_k, |\mathcal{E}_k|))$ per layer, using the selection probability sampling performed as follows:
\begin{align}
    \textbf{P}(i) = \frac{{(r_i)^{\gamma}}}{\sum_{j \in \mathcal{E}_k} {(r_j)^{\gamma}}}, \quad j\in\mathcal{E}_k
\end{align}
where $\gamma \in (0,1]$ is a decay exponent of reputation coefficient. Then, $\mathcal{M}_k$ is obtained by randomly selecting $m_k$ nodes in $\mathcal{E}_k$ based on selection probabilities $\textbf{P}(i)$ without replacement. Hence, it follows a hypergeometric distribution in probability theory and statistics~\cite{hald1960compound}, which can be formally stated as $\mathcal{M}_k\thicksim \operatorname{Hypergeometric}(m_k, \mathcal{E}_k, \textbf{P}(i))$. 


In the case of $\sum^L_{k=1}|\mathcal{M}_k|<\mathcal{K}$, we calculate a remaining quota $\mathcal{K}_{\text{remain}} = \mathcal{K}-\sum^L_{k=1}|\mathcal{M}_k|$ and gather all eligible nodes not yet selected as a global candidate pool, $\mathcal{E}_{\text{global}} = (\cup_{k=1}^L\mathcal{E}_k) \backslash (\cup_{k=1}^L\mathcal{M}_k)$. 
Again, using the selection probability sampling on the global eligible pool to calculate the remaining committee $\mathcal{M}_{\text{remain}}$, which is given by,
\begin{align}
    \textbf{P}'(i) = \frac{{(r_i)^{\gamma}}}{\sum_{j \in \mathcal{E}_{\text{global}}} {(r_j)^{\gamma}}}, \quad j\in\mathcal{E}_{\text{global}}
\end{align}
Hence, $\mathcal{M}_{\text{remain}} \thicksim \operatorname{Hypergeometric}(\mathcal{K}_{\text{remain}}, \mathcal{E}_{\text{global}}, \textbf{P}'(i))$, and the total committee is obtained by $\mathcal{M} = (\cup_{k=1}^L\mathcal{M}_k) \cup \mathcal{M}_{\text{remain}}$.

After the final selection, the system updates each node's statuses, including the selected history and the cooldown time $cd_i$, which is formulated by
\begin{equation}
    cd^{(t+1)}_i = \begin{cases}
            3, \quad \quad \quad \quad \quad \quad \quad \quad \quad \text{if }i\in \mathcal{M},\\
            \max(0, cd_i^{(t)}-1), \quad \quad  \text{otherwise}.
        \end{cases}
\end{equation}
where we specify that if $i$ is selected as a committee, the system will update the cooldown as three at the current round and decrease the cooling time of the unselected nodes by one cooldown per round. 

\vskip 0.1cm 
\noindent \textbf{Reputation update model.}
We employ a dynamic reputation update model to encourage long-term high-quality and stable contributions for all participants through a bonus for the quality and stability of contributions while discouraging malicious behaviours by penalties on reputation. 
To achieve that, we consider a dynamic decay factor $\delta\in[0,1]$ to incentivize long-term participation by obtaining a reputation compensation from the last round, which is given by  
\begin{align}
    \delta = \delta_b + \lambda_p\cdot(1-\frac{1}{1+\text{participation}_i/100}),
\end{align}
where $\delta_b$ is a base decay factor. $\lambda_p$ denotes a decay compensation parameter and $\text{participation}_i$ is the $i^{th}$ participant's historical participation times. As such, the more $\text{participation}_i$, the higher the reputation compensation $\delta \cdot r_i^{(t-1)}$ for round $t$. 

Besides, a good quality of contribution is good for an increased reputation. Let $X_c$ denote a unit bonus from the contribution quality to reputation updates, the amount of increased reputation resulting from good contribution is given by $X_c \cdot q_i^{(t)}$, where $q_i^{(t)}$ is quantified by Eq.~\ref{eq:vc}. 
Additionally, we define $\lambda_{\text{stab}} = 1 - {\operatorname{std}\bigl( [C_i^{t-\tau}, \cdots,C_i^{t}}] \bigr)/{\tau}$
as a stability parameter during the last $\tau$ rounds to avoid malicious nodes with random attacks to obtain rewards.
Hence, the new reputation at round $t$ can be defined by 
\begin{align}\label{eq:reputationupdates}
    r_i^{(t)} = \delta r_i^{(t-1)}+ q_i^{(t)} X_c + \lambda_{\text{stab}} X_s,
\end{align}
where $X_s$ denote unit bonus from the stability to reputation updates and $r_i^{(t)} \in [0, r_{\text{max}}]$. We employ a dynamic maximum capability $r_{\text{max}}$ per round to limit the upper bound of reputation updates.

\vskip 0.1cm
\noindent\textbf{Malicious detection and penalty.}
To safeguard the FL process, we consider malicious detection and penalty through a set of conditions. It consists of: 1) persistent low contributions (condition 1), 2) abnormal fluctuations (condition 2), and 3) sudden behavioural changes (condition 3), which serve as indicators for detecting malicious nodes. For ease of analysis, we assume that any participants' behaviours meet the condition set (condition 1 AND condition 2) OR condition 3, the system will detect them as a set of malicious nodes $N_{\text{malicious}}$ and perform a penalty to them. The penalty function is defined by, 
\begin{align}\label{eq:penalty}
    \text{penalty} = \min(\lambda_{r} r_j + \lambda_{s}  S_j, \frac{r_j}{2}),
\end{align}
where $\lambda_{r}, \lambda_{s}$ denote the factors of applying penalty to reputation updates and stake, respectively. The upper bound $\frac{r_j}{2}$ is intended to avoid excessive penalties on the reputation. 
Hence, the updated reputation for node $j$ at round $t$ is given by $r_j^{(t)} = r_j^{(t-1)}-\text{penalty}, \forall j\in N_{\text{malicious}}$.

\vskip 0.1cm
\noindent \textbf{Reward calculation.}
Let $B$ denote the base reward of the reward pool. The reward calculation considers a participant's stake, historical contribution performance, and committee bonus. The stake and contribution proportions relative to the overall records influence the rewards participants receive from the reward pool.

We introduce two dynamic weights, $\alpha$ and $\beta$, based on reputation to balance the influence of stake and historical contributions in reward calculation, where $\beta = 1 - \alpha$. The weight $\alpha$ is defined as $\alpha=\sigma(\frac{\overline{r}-r^0_{i}}{f_{\text{scale}}})\cdot \lambda_{\text{stake}}$, where $\lambda_{\text{stake}}$ is the global stake weight, and reputation $r$ determines the adjustment. 

To prevent monopolization, we introduce an effective stake limit for large nodes, ensuring no single node dominates. The effective stake is given by $S_i^{\text{eff}} = \min(S_i, 3\cdot \overline{S})$, where nodes holding more than three times the average stake $\overline{S}$ are capped.

The historical contribution of participant $i$ over the last $\tau$ rounds is computed as $C^{\text{hist}}_i=\sum_{t=0}^{\tau}C_i^{(t-\tau)}\cdot \zeta^{t}$, where $\zeta^t (\zeta \in (0,1))$ is an exponential decay factor that prioritizes more recent contributions. The total historical contribution across all participants is given by $C_{\text{total}}=\sum_{i\in N}\sum_{t=0}^{\tau}C_i^{(\tau-t)}\cdot \zeta^{t}$. 

Any participant selected as a committee member will receive a bonus for their contributions to model validation and aggregation, which is formulated by,
\begin{align}\label{eq:Rcmm}
    R_{\text{cmm}} = \begin{cases}
            B_{\text{cmm}}\cdot J_{\text{c}}(r), \quad \text{ if }i \in \mathcal{M}^{(t)},\\
            0, \quad \quad \quad \quad \quad \quad \text{otherwise.}
        \end{cases}
\end{align}
Here, $J_{\text{c}}(r) = \frac{(\sum_{i=1}^{\mathcal{K}} r_i)^2}{\mathcal{K}\cdot \sum_{i=1}^{\mathcal{K}} {r_i}^2 + \epsilon}\cdot \sigma(\frac{\overline{r}}{10})$, $B_{\text{cmm}}$ represents a base bonus, and $J_c(r)$  is an improved Jain’s fairness index~\cite{jain1984quantitative} calculated from the reputation scores of the committee. A small constant $\epsilon$ is introduced to prevent division by zero errors. The sigmoid function $\sigma(\frac{\overline{r}}{10})$  ensures that the average reputation score is mapped to the range $(0,1)$. When the mean reputation $\overline{r}$ is low, the overall score decreases accordingly. If the fairness index $J_c(r)$ is low, the committee reward $R_{\text{cmm}}$ is also reduced, which promotes fair and balanced incentives.

Furthermore, we employ $J(r)$ based on reputation, as a fairness indicator to ensure equitable reward distribution among all participants $i\in N$ when earning rewards from their stake and historical contributions. Consequently, the total reward allocated to each participant at round $t$ is formally defined as
\begin{align}\label{eq:reward}
    R_i^{(t)} = (\alpha \cdot B \frac{S_i^{\text{eff}}}{\sum S_j} +\beta \cdot B \frac{C^{\text{hist}}_i}{C_{\text{total}}})\cdot J(r) + R_{\text{cmm}}
\end{align}
Note that if a participant has no contribution history (i.e., $C_i^{\text{hist}}=\emptyset$ or $C_i^{(t)}=0$), the current reward is set to zero, $R_i^{(t)}=0$. 
Therefore, it enables adjustable weights and zero-contribution handling to calculate the participant's reward. The entire procedure can be implemented as a reputation-based consensus in a smart contract for reliable FL processes, whose pseudocode is presented in Algorithm~\ref{alg:reliableFL}.

\begin{algorithm}[!thb]
\DontPrintSemicolon
\KwOutput{Well-trained model $\mathbf{w}$, updated reputation $r$, reward $R$.}
\KwInput{FL workflow $F$, consisting of: an initial model $\mathbf{w}$, a set of $N$ participants indexed by $i$, an aggregation strategy $A$, and system configuration.}
    
    Initialize the FL workflow $F$;


    \tcc{Define the data structures:}

    struct \textbf{Node} \{
        id, stake, reputation, totalReward, violations, participation, cooldown, ..., identityVerified
    \}

    struct \textbf{SystemConfig} \{
        baseReward, committeeSize, stakeWeight, ..., maliciousPercent
    \}

    contract \textbf{FLSystem} \{ \tcp*{Define the main contract}

        $\quad$Initialize the nodes;

        $\quad$Initialize the SystemConfig;

        $\cdots$
        
        \tcc{Core logic of the main contract:} 
        $\quad$function \textbf{runRound}() \{ 
        
            $\quad$\For{each round $t$}{
            
                $\quad$ Collect $C^{(t)} = \{C_i^{(t)} \mid \forall i \in N\}$ within $T_{\max}$;

                $\quad\mathcal{M}^{(t)} \gets$ \text{selectCommittee}($r^{(t)}$, $\mathcal{K}$, $L$); 

                $\quad$committee performs aggregation;

                $\quad$\tcc{Malicious detection and penalty:}
                $\quad$\If{$\forall i \in N_{\text{malicious}} = \emptyset$}
                {
                    $\quad$Update reputation: $r^{(t)}_i \gets $ \textbf{Eq.}~\ref{eq:reputationupdates};
                }
                $\quad$\Else
                {
                    $\quad$Apply penalty: $penalty_i \gets$ \textbf{Eq.}~\ref{eq:penalty};
    
                    $\quad r^{(t)}_i \gets r^{(t-1)}_i - penalty_i,\forall i \in N_{\text{malicious}}$;
                    $\quad$$\text{violation}++$;
                }

                $\quad$Calculate rewards: $R_i^{(t)} \gets$ \textbf{Eq.}~\ref{eq:reward}, $\forall i \in N$;

                $\quad$$t++$; 

            }
            \tcc{Stop when an end condition triggers}
            $\quad$ \Return final model weights $\mathbf{w}$, $r$, $R$.
            
        $\quad$\}

    \textbf{Each FL client executes:} 

    $\quad$\For{each client $i \in N$ at round $t$ \textbf{in parallel}}  
    {
        $\quad$Update model and calculate contributions: $\mathbf{w}^{(t)}_i, C^{(t)}_i \gets$ \textbf{Eqs.}~\ref{eq:FLloss}, ~\ref{eq:contrib};
    }

    \textbf{Committee executes:} 

    $\quad$Aggregate model weights: $\mathbf{w}^{(t+1)} \gets$ \textbf{Eq.}~\ref{eq:FMA};

    \caption{Reputation-based reliable FL contract.}\label{alg:reliableFL}
\end{algorithm}

\vskip 0.1cm
\noindent\textbf{Contract optimality.}
With the above formulated optimization problem and detailed reward model, we can now study the optimality of the contract problem.

Let $\phi^*_i \triangleq (C^*_i, S^*_i, R^*_i)\in \Phi^*, \forall i\in N$ in the new ordering be the optimal contract that the task publisher can derive from maximizing its total profit $\Pi(\cdot)$ and each participant performs optimal behaviours to maximize its utility. These contract items are found through solving the optimization problem as presented in Eq.~\ref{eq:incentive problem} under the IC and IR constraints.
To derive the optimal contract for this problem, there may be several solutions. 


For ease of analysis, we simplify the derivation process. By assuming that: 1) all participants comply with the rules viz $\mathbb{I}_{\text{compl}}=1$ (no violations), 2) consider a single participant scenario that can ignore the IC constraint, and 3) the cost of contribution $c(C) = \frac{1}{2}\gamma_cC^2$ where $\gamma_c>0$ is a cost parameter and $C\in [0, C_{\max}]$, the relaxation version of the optimal problem can be formulated as
\begin{align}
    \max_{C,S,R} \Pi = V(C) - R, \quad 
    \text{s.t. } R - \frac{1}{2}\gamma_cC^2 \geq 0.
\end{align}
If we ignore the IR constraint and directly maximize profit $\Pi$, we need to define the relationship between the reward $R$ (Eq.~\ref{eq:reward}) and contribution value $V(C)$ (Eq.~\ref{eq:vc}). We further assume 1) there are no monopoly stakes $S\leq 3\overline{S}$, then $S^{\text{eff}}=S$, 2) $r_i = \overline{r}$, then $\alpha=\lambda_{\text{stake}}$, and 3) both $C^{\text{hist}}$ and $C_{\text{total}}$ are constants. 
At this point, maximizing $\Pi$ is equivalent to differentiating the objective function with respect to the first term $C$ and solving the first-order condition, which can be rearranged as 
\begin{align}
    \frac{\partial\Pi}{\partial C} =  V'(C) - &\frac{(1-\lambda_{\text{stake}})BJ(r)}{C_{\text{total}}}\cdot\frac{\partial C^{\text{hist}}}{\partial C} =0,
\end{align}
where we assume $C^{\text{hist}} = C\cdot\sum_{t=0}^{\tau}\zeta^{t}$, which leads to
\begin{align}\label{eq:v'c1}
    V'(C) = \frac{(1-\lambda_{\text{stake}})BJ(r)}{C_{\text{total}}}\cdot\frac{1-\zeta^{\tau+1}}{1-\zeta}.
\end{align}

Correspondingly, we define $k = \frac{1}{C_{\max}}$ to extend the $V(C)$ as the full version and calculate its derivative $V'(C)$, which is calculated by,
\begin{align}\label{eq:v'c2}
    V'(C) =\frac{\partial(\frac{X_c}{\tau}\cdot\frac{1}{1+e^{-kC}})}{\partial C}  = \frac{X_c\cdot ke^{-kC}}{\tau(1+e^{-kC})^2}, 
\end{align}
Thus, the optimal contribution $C^*$ made by participant can be derived from $V'(C^*)$ (Eq.~\ref{eq:v'c1}) = $V'(C^*)$ (Eq.~\ref{eq:v'c2}), which is given by
\begin{align}
    C^* = \frac{1}{k}\ln(\frac{X_c\cdot k\tau(1-\zeta)}{(1-\lambda_{\text{stake}})BJ(r)(1-\zeta^{\tau+1})C_{\text{total}}}-1).
\end{align}

Since $\frac{1}{k}=C_{\max}$, then we have $C^* = C_{\max}\cdot y$ where $y=\ln(x-1)$ is a logarithmically increasing function for $x=\frac{X_c\cdot k\tau(1-\zeta)}{(1-\lambda_{\text{stake}})BJ(r)(1-\zeta^{\tau+1})C_{\text{total}}}$ that asymptotically converges, theoretically, the optimal contribution would approach to positive infinity. However, the maximum practical constraints $C\in [C_{\min}=0, C_{\max}]$, as a result, the actual optimal solution is $C^* = C_{\max}$. In other words, $C_i^*\propto f_{\text{local}}(D_i, \mathbf{w})\cdot g_{\text{behaviour}}(h_i)$, adhering to optimal local training while demonstrating good behaviour, becomes the dominant strategy to maximize its utility for any participant. Meanwhile, $\ln(\frac{X_c\cdot k\tau(1-\zeta)}{(1-\lambda_{\text{stake}})BJ(r)(1-\zeta^{\tau+1})C_{\text{total}}}-1)=1$ can serve as a guideline for tuning the parameters (e.g., $\lambda_{\text{stake}}$, $X_c$ and $B$ if given fixed others) to achieve the optimal solution. 
By binding IR constraint, the optimal reward is $R^* = c(C^*)$. As all participants stake the same amount $S$, which leads to
\begin{align}
    S^* = &\frac{\lambda_{\text{stake}}BJ(r)}{n\cdot(R^*-R_{\text{cmm}}-\frac{(1-\lambda_{\text{stake}})BJ(r)C^{\text{hist}}}{C_{\text{total}}})}, \\
    R^* = &c(C^*).
\end{align}

If we consider the IR constraint, the optimization problem can be reorganized as a Lagrangian function with a Lagrange multiplier and then repeat the derivation process for the three terms $C, S, R$, respectively. With the help of math tools (e.g., SciPy methods in Python) to solve for the problem, we can obtain the optimal $C^*, S^*, R^*$ by adjusting the parameters as needed by numerical analysis. Suppose the distribution of the participants' violations (concerning Byzantine failures) is not predictable and the stakes are unbalanced. In those cases, we can consider using advanced approaches, such as Bayesian priors~\cite{frazier2018tutorial} or reinforcement learning optimization~\cite{kaelbling1996reinforcement}, to meet the requirements.

\section{Experiments and Evaluation}\label{sec:reliableFL-evaluation}

This section illustrates simulation setup, demonstrates the feasibility of the proposed incentive mechanisms, and evaluates the outcomes for validation and guidance of (optimal) smart contract design. 

\subsection{Simulation Setup}

In the simulation, we conducted extensive simulation experiments on a local computer equipped with an Apple M1 chip with eight (four performance and four efficiency) cores, 16GB memory, and a 500 GB Macintosh HD disk. 

For the evaluation of solutions, we utilize a normal distribution $\mathcal{N}$ with random fluctuation $\mathcal{F}$ for the ease of training process to present the normal behaviour pattern for participants acting normally, $C_{\text{normal}} = \max(0, \mathcal{N}(\mu=7, \sigma^2=1)\cdot \mathcal{F})$, 
where $\mu$ and $\sigma$ denote the mean and standard deviation of the set of contributions collected from the normal (or honest) FL participants. 
For malicious behaviour patterns, we simulate malicious patterns with three different attacks: 1) false high contribution attack, where malicious nodes disguise themselves as high-value nodes through abnormally high contributions to avoid detection based on low contributions; 2) zero contribution attack where malicious nodes directly destroy the federated model aggregation and reduce model performance, and 3) random attack which is given by a random choice of 60\% false high contribution attack and 40\% zero contribution attack. 

We introduce a parameter $\eta_{\text{switch}}$ that allows switching between different patterns to simulate various malicious behaviours during the training process. The parameter defines the longest time window threshold at which the system transitions from base to progressive malicious phases --- ranging from high-contribution and zero-contribution attacks to randomly hybrid attacks. This enables the implementation of malicious pattern-switch logic for progressive attack testing, periodic behavioural perturbations, and defense mechanism verification. By periodically switching attack strategies (e.g., every 30 rounds), we simulate the adaptability of malicious participants. Additional details are available in the source code.

Additionally, we set different malicious percentages as an adjustable parameter to distinguish between malicious and honest nodes among the total participants. An attack detection function is implemented to identify such malicious behaviours, which allows us to test and evaluate the effects of penalties, reputation updates, and rewards on the system. A summary of the other parameters used in the simulation is presented in Table~\ref{tab:para}.

\begin{table}[!htb]
    \centering
    \caption{Parameter setting in the simulation.}
    \label{tab:para}
    {
    \resizebox{0.48\textwidth}{!}{
    \begin{tabular}{ccc}
         \toprule
         \textbf{Parameter} & \textbf{Description}& \textbf{Setting} \\
        \midrule
        $S_i$ & initial stake & 100 \\
        $r_i^0$ & initial reputation score & 100 \\
        $C_{\min}$ & minimal contribution score & 0 \\
        $C_{\max}$ & maximum contribution score & 10 \\
        $\gamma$ & reputation decay coefficient & 0.5 \\
        $r^{t\leq5}_{\max}, r^{t>5}_{\max}$ & maximum capabilities of reputation updates & 300, 500 \\
        $\epsilon$ & a small constant & $10^{-8}$ \\
        $cd(\cdot)$	& cooldown period	& 3 \\
        $L$ & number of strata & 3\\
        $B_{\text{cmm}}$	& base bonus to committee member & 40 \\
        $\delta_b$	& base decay factor & 0.88 \\
        $\lambda_p$	& decay compensation parameter	& 0.07 \\
        $\tau$	& recent historical rounds	& 5 \\
        $\tau_{\text{stab}}$ & default stability for new participants & 0.8 \\
        $X_c$ & base bonus of contribution & 50 \\
        $X_s$ & base bonus of stability & 30 \\
        $\lambda_r$	& penalty factor to reputation	& 0.3 \\
        $\lambda_s$ & penalty factor to stake & 0.1 \\
        $\zeta$ & historical decay factor & 0.9 \\
        $B$ & base reward value in the pool & 1200 \\
        $n$ & the number of participants & 100 \\
        $\lambda_{\text{stake}}$ & stake weight & 0.4 \\
        $\mathcal{K}$ & committee size &  5 \\
        $m$ & malicious percent & 15\%\\
        $\eta_{\text{switch}}$ & the first time window of observing a switch of malicious patterns & 5\\
        $Round$ & the total number of the FL rounds & 90\\
        \bottomrule
    \end{tabular}
    }
}
\end{table}

To measure the fairness in reward allocation based on actual contributions, we consider two metrics for evaluation. Jain's fairness index~\cite{jain1984quantitative} is a quantitative measure of fairness and discrimination for resource allocation in shared systems. We have used it to maintain the fairness of committee selection on reputation presented in Eq.~\ref{eq:Rcmm}, whereas it can also be employed to measure the fairness of reward allocation $J(R)$ by replacing the input as $R$. Additionally, we consider an opposite metric --- the Gini coefficient~\citep{dorfman1979formula}, as a measurement of reward inequality, which is formulated by
\begin{align}\label{eq:fair&gini}
    J(R) = \frac{(\sum_{i=1}^n R_i)^2}{n\sum_{i=1}^n {R_i}^2 + \epsilon}\cdot \sigma(\frac{\overline{R}}{10}),\\
    G (R) = (n+1-2\frac{\sum_{i=1} ^n\sum_{j=1}^iR_j}{\sum_{j=1}^n R_j})/{n}    
\end{align}
where the input set $R$ is performed in ascending order. The source code is available online at the GitHub repository\footnote{\url{https://github.com/yuandou168/reliableFLOps}}.

\subsection{Simulation Results}
\begin{figure}[!htb]
    \centering
    \subfloat[Total malicious nodes detected per round.]
    { 
    \label{subfig:N100_M15_p1}
      \includegraphics[width=.96\linewidth]{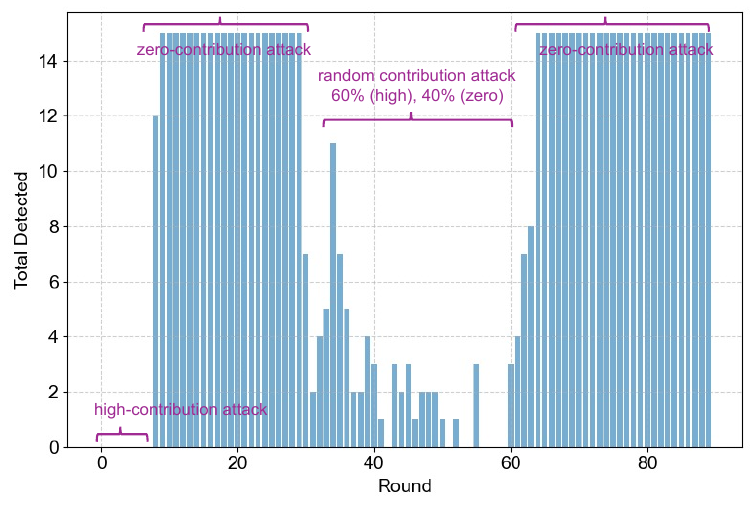}} \\
    \subfloat[Reputation update dynamics.]
    { 
    \label{subfig:N100_M15_p2}
      \includegraphics[width=.96\linewidth]{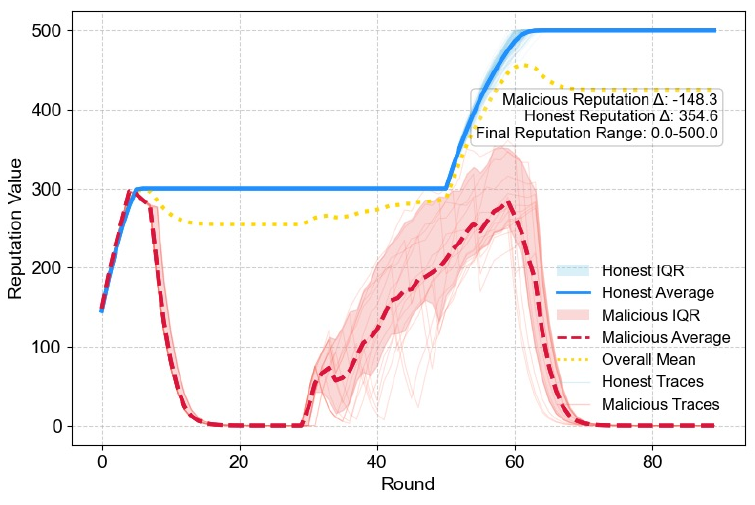}} \\
    \subfloat[Reward distribution.]
    { 
    \label{subfig:N100_M15_p3}
      \includegraphics[width=.96\linewidth]{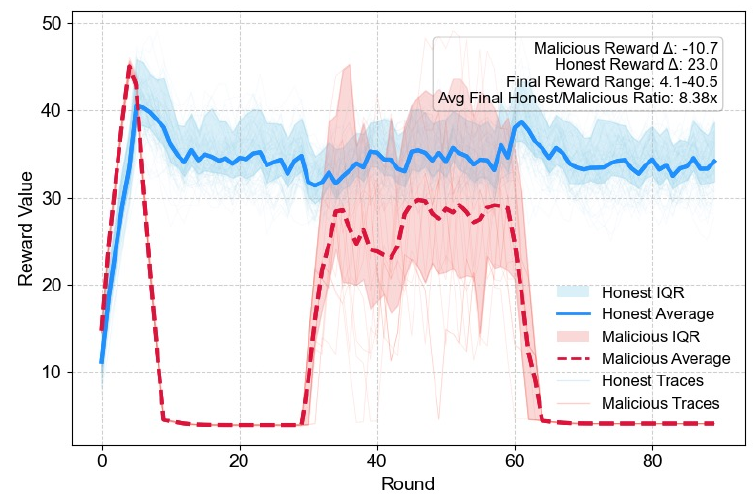}} \\
    \caption{Overview of malicious detection, reputation updates, and reward distribution per round.}
    \label{fig:switch-phase-round=5}
\end{figure}

This section presents an overview of the simulation results under the parameter setting presented in Table~\ref{tab:para}. Figure~\ref{fig:switch-phase-round=5} showcases the dynamics of malicious node detection, reputation updates, and reward distribution per round. 

\vskip 0.1cm 
\noindent \textbf{Malicious detection.} Sub Figure~\ref{subfig:N100_M15_p1} illustrates the malicious detection performance per round. The detection method follows the condition set \textit{(condition 1 AND condition 2) OR condition 3}, based on a malicious percentage of $15\%$. The malicious pattern switch parameter $\eta_{\text{switch}}$ plays a crucial role in shaping the behaviour records of malicious participants, thereby influencing detection performance.

Notably, false high-contribution malicious nodes were not detected during the first five rounds, as the current detection conditions do not account for high-contribution attacks, allowing such behaviours to appear normal. When the round $t \in [5, 30)$, the malicious pattern shifts to zero contributions. By the eighth round, the system successfully identified 12 out of 15 malicious nodes. When the round $t \in [30, 60)$, the malicious pattern transitions to random attacks, with a 60\% probability of high-contribution attacks and a 40\% probability of zero-contribution attacks. In the final 30 rounds, malicious behaviours revert to zero-contribution attacks. The detection method proves to be more effective against zero-contribution attacks but remains ineffective against false high-contribution attacks due to current detection limitations.

\vskip 0.1cm
\noindent \textbf{Reputation updates.} Sub Figure~\ref{subfig:N100_M15_p2} illustrates the reputation update dynamics per round, distinguishing between honest and malicious nodes among 85 honest participants and 15 malicious ones. Initially, all nodes start with a reputation score of 100, which increases as new contributions are collected during the first few rounds. This growth continues until it reaches a plateau around a reputation score of 300. Notably, there is no difference in reputation growth between honest and malicious nodes until the eighth round.

At the eighth round, some malicious nodes experience a sharp drop in reputation due to detected malicious behaviours, as shown in Figure~\ref{subfig:N100_M15_p1}. Since most malicious nodes were identified during this period, it caused their reputation scores to decline significantly, eventually approaching zero. During the random attack phase, the system fails to detect all malicious activities, allowing some malicious nodes to temporarily regain their reputation through high-contribution attacks. This results in fluctuations in their reputation updates, as not all malicious nodes receive penalties in this phase. However, once all 15 malicious nodes are successfully detected, the fluctuations cease, and their reputation scores rapidly decline, dropping by an average of 148.3 points.

Meanwhile, the reputation of the 85 honest nodes steadily increased as expected in the first five rounds by 300. Then they gradually reach the upper limit of 500. The average reputation increase for honest nodes is approximately 354.6, significantly higher than that of the malicious nodes.


\vskip 0.1cm
\noindent \textbf{Reward distribution dynamics.} Sub Figure~\ref{subfig:N100_M15_p3} illustrates the reward distribution dynamics among 85 honest and 15 malicious nodes out of 100 participants. Because malicious nodes mimic high-contribution participants, they accumulate higher rewards than honest nodes. Hence, they exhibit a higher average reward value and a faster growth rate in the early rounds. However, as the system detects zero-contribution attacks, their rewards drop sharply to zero. Although some malicious nodes temporarily regain their reputation by engaging in high-contribution attacks in the period of random attacks, as more malicious nodes are progressively detected, their corresponding rewards decline. 

Although all malicious nodes ultimately reach a very low reward level over the 90 rounds, their total rewards are not zero. Overall, the final reward values range between 4.1 and 40.5 at round 90, satisfying the IR constraint. However, participants acting maliciously or continuing such behaviour under this reward incentive mechanism cannot achieve optimal utility. Specifically, the average reward for malicious nodes dropped by 10.7, while the average reward for honest nodes increased by 23.0. The total reward accumulated by the 85 honest nodes is 8.38 times greater than that of the 15 malicious nodes in the simulation system. 

Consequently, this incentive mechanism encourages participants to adopt a more rational strategy --- avoiding malicious behaviour to maximize their rewards.
Therefore, it preserves the trend of rewarding aligned with the incentives while reducing the risk of over-penalty and over-reward. 

\begin{figure}[!htb]
    \centering
    \includegraphics[width=.96\linewidth]{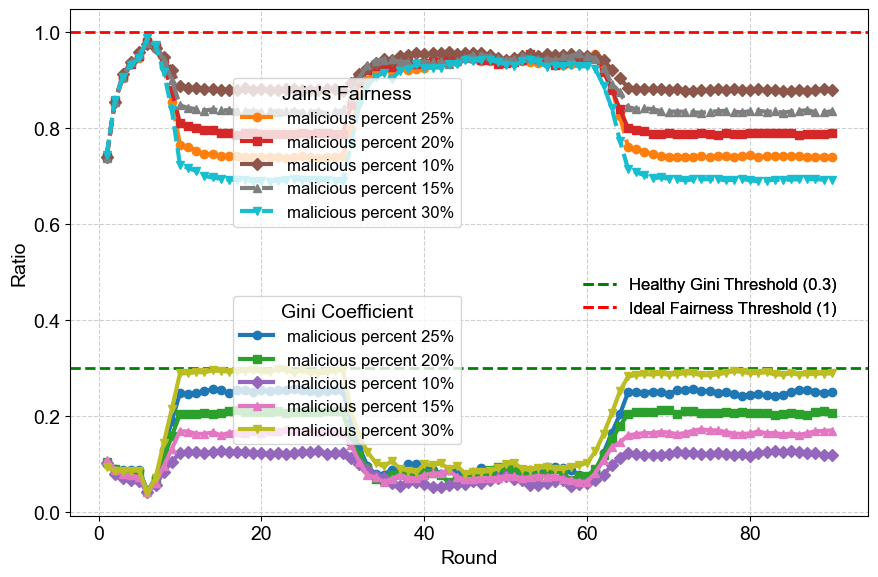}
    \caption{Comparisons of total fairness metric dynamics across the last 90 rounds.} \label{fig:fairness}
\end{figure}

\vskip 0.1cm
\noindent \textbf{Fairness.} 
Figure~\ref{fig:fairness} compares the fairness of reward distribution across 100 participant nodes under five different malicious percentage settings. Similar to the reward dynamics observed in Figure~\ref{subfig:N100_M15_p3}, the fairness initially improves, as indicated by an increase in Jain's fairness index and a decrease in the Gini coefficient (see Eq.~\ref{eq:fair&gini}), that suggests a trend toward a more equitable reward distribution. However, due to the influence of malicious nodes, the reward allocation is not always fair. 


For instance, given a malicious percentage 15\%, we observed that malicious detection significantly influences reward allocation's fairness. When most or all malicious nodes are successfully identified, i.e., during rounds between 8 and 30, as well as 60 to 90, the fairness index decreases while the Gini coefficient increases. Conversely, fairness remains relatively stable when the system detects only a few malicious nodes. It reflects an imbalanced reward allocation influenced by the performance of malicious detection.

As the percentage of malicious nodes in the system increases, e.g., from 10\% to 30\%, the reward distribution becomes more inequitable between honest and malicious nodes. However, such inequity remains at a healthy threshold (<0.3). Such a change is positive because it still preserves fairness among honest participants while preventing malicious participants from obtaining unfair rewards.

\vskip 0.1cm
\noindent \textbf{The optimal parameter space.} In the simulation, we employ the controlled variable method and an optimization solver `SLSQP', which is suitable for constrained optimization, to analyze the applicable approximations of the optimal contract items $(C_i^*, S_i^*, R_i^*)\in\Phi^*$. The numerical analysis validates the optimal contribution $C^* \rightarrow C_{\max}$ for all participants. It also verifies that the IR-constrained satisfaction rate is 100\% while minimum utility is more than zero. With this numerical analysis, users can adjust their contract terms for different requirements with optimal parameter settings. More technical details have been presented in the source code.  

These simulation results suggest that the proposed incentive mechanisms are feasible in an FL system with honest and malicious participants. By coding these mechanisms into smart contracts and bounding blockchain, the system can effectively constrain and guide participant behaviour, ensuring that honest participants can receive fair compensation while discouraging malicious activities. The numerical analysis can help customize (approximately) optimal contract items written in smart contracts to adopt diverse FL scenarios.

\section{Summary}\label{sec:reliableFL-conclusion}
This study demonstrated how FL can be structured as a workflow using open workflow standards and executed on remote infrastructure to address automation challenges. To bridge the gap between traditional workflow-based automation and decentralized collaboration, we propose a novel strategy that builds upon our prior work to enhance reliability in FL management over decentralized infrastructures. We leverage contract theory with a multidimensional scheme to tackle the incentivizing collaboration problem by constructing a contribution model, implementing fair committee selection, dynamically updating reputations, calculating rewards, and defining corresponding profit and utility functions. Additionally, we explore the optimality of contracts to guide the design and implementation of smart contracts that can be deployed on blockchain networks to assure training quality.
We conduct extensive simulation experiments to validate the proposed approach. The results demonstrate that our incentive mechanisms are effective, ensuring fair reward allocation despite malicious attacks. Future improvements include encoding the optimal contract items into smart contracts for real-world system performance evaluation and prototyping an integrated on-chain and off-chain system demonstration.

\section*{Acknowledgments}
We thank Mr. Anandan Krishnasamy for running FL workflow experiments over the hybrid cloud environment and validating the FL training results. This research was made possible through partial funding from several European Union projects: CLARIFY (860627), ENVRI-Hub Next (101131141), EVERSE (101129744), BlueCloud-2026 (101094227), OSCARS (101129751), LifeWatch ERIC, BioDT (101057437, through LifeWatch ERIC), and Dutch NWO LTER-LIFE project.

\bibliographystyle{unsrtnat}
\bibliography{refs}

\end{document}